\documentclass[runningheads]{llncs}
\usepackage{balance}
\usepackage{amsmath}
\usepackage[tight,footnotesize]{subfigure}
\usepackage{bm}
\usepackage{algorithm}
\usepackage[noend]{algorithmic}
\usepackage{multirow}
\usepackage{amsfonts}

\begin{document}




\title{Achieving Long-term Fairness in Submodular Maximization through Randomization}
\titlerunning{Achieving Fair Submodular Maximization through Randomization}
%
\author{Shaojie Tang\inst{1} \and Jing Yuan\inst{2} \and Twumasi Mensah-Boateng \inst{2}}
%
%
\institute{Naveen Jindal School of Management, University of Texas at Dallas\\\email{shaojie.tang@utdallas.edu} \and Department of Computer Science and Engineering, University of North Texas \\\email{jing.yuan@unt.edu} \email{twumasimensah-boateng@my.unt.edu}}

\maketitle              
\begin{abstract}Submodular function optimization has numerous applications in machine learning and data analysis, including data summarization which aims to identify a concise and diverse set of data points from a large dataset. It is important to implement fairness-aware algorithms when dealing with data items that may contain sensitive attributes like race or gender, to prevent biases that could lead to unequal representation of different groups. With this in mind, we investigate the problem of maximizing a monotone submodular function while meeting group fairness constraints. Unlike previous studies in this area, we allow for randomized solutions, with the objective being to calculate a distribution over feasible sets such that the expected number of items selected from each group is subject to constraints in the form of upper and lower thresholds, ensuring that the representation of each group remains balanced in the long term. Here a set is considered feasible if its size does not exceed a constant value of $b$. Our research includes the development of a series of approximation algorithms for this problem.
\end{abstract}


\section{Introduction}
A set function is referred to as submodular if it follows the principle of diminishing returns, where adding an item to a larger set yields a smaller benefit. This concept is applied in various real-world scenarios such as feature selection \cite{das2008algorithms}, where the goal is to select the most relevant features from a large pool of potential features to use in a machine learning model;  active learning \cite{golovin2011adaptive}, where the goal is to choose a set of instances for a machine learning model to learn from;  exemplar-based clustering \cite{dueck2007non}, where the goal is to choose a set of exemplars to represent a set of data points;  influence maximization in social networks \cite{tang2020influence}, where the goal is to choose a set of individuals to target in order to maximize the spread of information or influence in a network; as well as recommender system \cite{el2011beyond} and diverse data summarization \cite{sipos2012temporal}. The goal of submodular optimization is to choose a set of items that optimizes a submodular function while satisfying constraints such as size limitations, matroid requirements, or knapsack restrictions.

In practice, items or individuals are often grouped based on attributes such as gender, race, age, religion, or other factors. However, if not properly monitored, existing algorithms may display bias and result in an over- or under-representation of certain groups in the final selected set. To address this issue, we propose the study of long-term fair submodular maximization problem. The aim is to randomly choose a subset of items that optimizes a submodular function, such that the expected number of selected items from each group falls within the desired range. This approach ensures that the final selection of items is not only optimized, but also equitable, providing a fair representation of all groups in the long term.

Formally, we consider a set $V$ of items, which are divided into $m$ (not necessarily disjoint) groups: $V_1, V_2, \cdots, V_m$ with items in each group sharing similar attributes (e.g., race). To ensure fairness, a randomized item selection algorithm must satisfy the following criteria for all groups $t \in [m]$ where $[m]=\{1, 2, \cdots, m\}$: the \emph{expected} number of selected items from group $V_t$ must be within the range of $[\alpha_t, \beta_t]$, where $\alpha_t$ and $\beta_t$ are arbitrary parameters that may differ across groups; moreover,  the number of chosen items must \emph{always} stay below a cardinality constraint of $b$. To put it simply, a fair randomized solution must meet two important requirements \cite{bera2019fair}: (a) restricted dominance, which means the proportion of items from each group must be within a certain limit, and (b) minority protection, which means the proportion of items from each group must not fall below a certain limit. Our fairness notation has gained significant recognition in the academic world and it has been adopted  in various studies, including multi-winner voting systems \cite{celis2018multiwinner}, fair recommendation systems \cite{el2020fairness}, and matroid-constrained optimization problems \cite{chierichetti2019matroids}. In fact, this notation is capable of capturing other fairness definitions such as statistical parity \cite{10.1145/2090236.2090255}, the $80\%$-rule \cite{biddle2017adverse},  and proportional representation \cite{monroe1995fully}.

Different from the majority of prior studies that concentrate on finding a \emph{fixed} set of items that comply with fairness restrictions, our approach accommodates randomized solutions and therefore offers greater flexibility in fulfilling the fairness restrictions. Take fairness-aware product recommendations as an example. The objective is to suggest a set of products to online consumers while ensuring that each group of sellers, such as male and female sellers, is expected to have at least one of their products recommended. Due to limited display space, suppose we can only display one product to the consumer. In this scenario, it is not possible for any of the deterministic solutions to fulfill the fairness requirement, however, a randomized solution can be easily found to satisfy it. For instance, a product can be suggested from each group with the same likelihood of occurrence.
\subsection{Our Contributions}

\begin{itemize}
\item We are the first to investigate the long-term fair submodular maximization problem, which presents a substantial challenge due to its exponential number of variables. As a result, it is difficult to solve using traditional linear programming (LP) solvers.
\item We develop a $(1-1/e)^2$-approximation algorithm that approximately satisfies the fairness constraints. Specifically, our algorithm ensures that the number of selected items from group $V_t$ is within the range of $[\lfloor\alpha_t\rfloor, \lceil\beta_t\rceil]$. Notably, if both $\alpha_t$ and $\beta_t$ are integers, our solution strictly satisfies the fairness constraints in the original problem.
\item It is important to note that the previous algorithm requires optimizing a continuous approximation of the underlying submodular function, referred to as the multi-linear extension \cite{calinescu2007maximizing}. This is achieved by executing the continuous greedy algorithm, whose implementation is computationally expensive in practice. Our second contribution is the introduction of a fast greedy algorithm that achieves a degraded approximation ratio of $(1-1/e)^2/2$.
\item We present a $(1-1/e)$-approximation randomized algorithm. Our approach involves utilizing the ellipsoid method and incorporating an approximate separation oracle for the dual LP of the original problem, which has a polynomial number of variables and an exponential number of constraints. Unlike the deterministic solutions, our randomized approach provides three key benefits. Firstly, our solution does not depend on the assumption of non-overlapping groups. Secondly, our approach strictly satisfies all fairness constraints. Thirdly, we achieve the optimal approximation ratio of $1-1/e$.

\end{itemize}

\subsection{Additional Related Works}
The growing recognition of the importance of fair and objective decision-making systems has resulted in a surge of interest in developing fair algorithms in various fields such as influence maximization \cite{tsang2019group} and classification \cite{zafar2017fairness}. The development of fair algorithms has also been applied to voting systems\cite{celis2018multiwinner}, where the goal is to ensure that election outcomes are a fair representation of the preferences of the voters. Moreover, the field of bandit learning \cite{joseph2016fairness}, which involves making sequential decisions based on uncertain information, has also seen a growing interest in the development of fair algorithms. Finally, the field of data summarization \cite{celis2018fair} has seen a growing focus on the development of fair algorithms, which aim to provide a balanced representation of the data.  The specific context and type of bias being addressed influence the choice of fairness metric adopted in existing studies, leading to various optimization problems and fair algorithms tailored to the specific requirements of each application. Our definition of fairness is broad enough to encompass many existing notations, such as the $80\%$-rule \cite{biddle2017adverse}, statistical parity \cite{10.1145/2090236.2090255}, and proportional representation \cite{monroe1995fully}. Unlike the majority of previous research on fairness-aware algorithm design \cite{celis2018multiwinner,el2020fairness,chierichetti2019matroids,tang2022group,yuan2023group}, which aims to find a deterministic solution set, our goal is to compute a randomized solution that can meet the group fairness constraints on average.

\section{Preliminaries and Problem Statement}

A set $V$ of $n$ items is considered and there is a non-negative submodular utility function $f: 2^V\rightarrow \mathbb{R}_+$. The marginal utility of an item $e\in V$ on a set $S\subseteq V$ is denoted as $f(e\mid S)$, i.e., $f(e\mid S)=f(\{e\}\cup S)-f(S)$. The function $f$ is considered submodular if, for any sets $X, Y\subseteq V$ with $X\subseteq Y$ and any item $e \in V\setminus Y$, the following inequality holds:
\[f(e\mid Y) \leq f(e\mid X).\] It is considered monotone if, for any set $X\subseteq V$ and any item $e \in V\setminus X$, it holds that
\[f(e\mid X) \geq 0.\]

Assuming $V$ is divided into $m$ groups, $V_1, V_2,\cdots, V_m$, there is a specified lower and upper bound on the \emph{expected} number of items from each group that must be included in a feasible solution. These bounds, referred to as $\alpha\in \mathbb{R}_{\geq0}^{m}$ and $\beta\in \mathbb{R}_{\geq0}^{m}$, represent group fairness constraints. In addition, there is a \emph{hard} constraint $b$ on the number of selected items. Let $\mathcal{F}=\{S\subseteq V\mid |S|\leq b\}$ denote the set of feasible selections.  The goal of the fair submodular maximization problem (denoted as $\textbf{P.0}$) is to determine a distribution $x\in [0,1]^\mathcal{F}$ over sets from $\mathcal{F}$ that maximizes the expected utility, while ensuring that the expected number of items selected from each group meets the fairness constraints. I.e.,

 \begin{center}
\framebox[0.55\textwidth][c]{
\enspace
\begin{minipage}[t]{0.55\textwidth}
\small
$\textbf{P.0}$
$\max_{x\in [0,1]^\mathcal{F}} \sum_{S\in \mathcal{F}}x_S f(S)$ \\
\textbf{subject to:}  \\
\begin{equation*}
\begin{cases}
\alpha_t\leq \sum_{S\in \mathcal{F}}(x_S\cdot |S \cap V_t|) \leq \beta_t, \forall t \in[m].\\
\sum_{S\in \mathcal{F}}x_S\leq 1.
\end{cases}
\end{equation*}
\end{minipage}
}
\end{center}
\vspace{0.1in}

Here each decision variable $x_S$ represents the selection probability of $S\in \mathcal{F}$. This LP has a total of $2m+1$ constraints, excluding the obvious constraints that specify that $x_S\geq 0$ for all $S\in \mathcal{F}$. Despite this, the number of variables in the LP problem is equal to the number of elements in $\mathcal{F}$, which can be exponential in $n$. As a result, conventional LP solvers are unable to solve this LP problem efficiently. The next lemma asserts that $\textbf{P.0}$ is a problem that is NP-hard.

\begin{lemma}
Problem $\textbf{P.0}$ is NP-hard.
\end{lemma}
\emph{Proof:} We demonstrate this by reducing it to the classic cardinality constrained monotone submodular maximization problem, which we will describe below.
\begin{definition} The cardinality constrained monotone submodular maximization problem takes as input a collection of items $V$, a monotone submodular function $f: 2^V\rightarrow \mathbb{R}_+$, and a cardinality constraint $b$. The goal is to choose a subset of items $S\subseteq V$ that maximizes $f(S)$ while ensuring that $|S|\leq b$.
\end{definition}

To show the reduction, we take an instance of the cardinality constrained monotone submodular maximization problem and create a corresponding instance of $\textbf{P.0}$. To do this, we consider only one group with no fairness constraints, meaning $V=V_1$, with $\alpha_1=0$ and $\beta_1= |V|$. It can be easily verified that the optimal solution of this instance is a distribution over a set of solutions, each of which is an optimal solution to the instance of cardinality constrained monotone submodular maximization problem. Additionally, although $\textbf{P.0}$ allows for randomized solutions, there exists at least one optimal solution that is a deterministic set. Specifically, every optimal solution of the cardinality constrained monotone submodular maximization problem must be an optimal solution to its corresponding instance of $\textbf{P.0}$. Hence, these two instances are equivalent. This concludes the proof of the reduction. $\Box$

\section{Near Feasible Deterministic Algorithms}
\label{sec:adaptive-general}
In this section, we present a deterministic algorithm for $\textbf{P.0}$. Here we assume that  $m$ groups do not overlap with each other. To begin, we introduce the multilinear extension of a monotone submodular function $f$. Given a vector $y\in [0,1]^n$, let $S_y$ be a random set where each item $i\in V$ is independently added to $S_y$ with probability $y_i$. Then we let
\[F(y)=\mathbb{E}[f(S_y)]=\sum_{S\subseteq V} f(S) \prod_{i\in S} y_i \prod_{i\notin S} (1-y_i).\]

We next introduce a new optimization problem $\textbf{P.1}$. The goal of $\textbf{P.1}$ is to compute a vector $y\in [0,1]^n$ that maximizes $ F(y)$ such that $\alpha_t\leq \sum_{i\in V_t} y_i \leq \beta_t, \forall t \in[m]$ and $\sum_{t\in [m]} \sum_{i\in V_t} y_i \leq b$.
 \begin{center}
\framebox[0.4\textwidth][c]{
\enspace
\begin{minipage}[t]{0.4\textwidth}
\small
$\textbf{P.1}$
$\max_{y\in [0,1]^n} F(y)$ \\
\textbf{subject to:}  \\
\begin{equation*}
\begin{cases}
\alpha_t\leq \sum_{i\in V_t} y_i \leq \beta_t, \forall t \in[m].\\
\sum_{t\in [m]} \sum_{i\in V_t} y_i \leq b.
\end{cases}
\end{equation*}
\end{minipage}
}
\end{center}
\vspace{0.1in}

The following lemma establishes a connection between the optimal solution of problem $\textbf{P.0}$ and that of problem $\textbf{P.1}$. This lemma serves as a crucial foundation for understanding the relationship between the two problems and allows for the development of a near optimal solution for $\textbf{P.0}$ by solving $\textbf{P.1}$.
\begin{lemma}
\label{lem:1}
Let $x^*$ denote the optimal solution of $\textbf{P.0}$ and $y^*$ denote the optimal solution of $\textbf{P.1}$, it holds that
\begin{eqnarray}
 (1-1/e)\sum_{S\in \mathcal{F}}x^*_S f(S) \leq F(y^*).
\end{eqnarray}
\end{lemma}
\emph{Proof:}  Let $B$ be a polytope defined as the set of all vectors $y\in [0,1]^n$ that meet the conditions in $\textbf{P.1}$, i.e.,
\begin{eqnarray}
\label{eq:2}
B=\{y\in [0,1]^n \mid \alpha_t\leq \sum_{i\in V_t} y_i \leq \beta_t, \forall t \in[m]; \sum_{t\in [m]} \sum_{i\in V_t} y_i \leq b; 0\leq y_i \leq 1, \forall i\in V\}.
 \end{eqnarray}
Given  the optimal solution $x^*$ of $\textbf{P.0}$, we then introduce a vector $\hat{y}\in [0,1]^n$ such that $\hat{y}_i=\sum_{S\in \mathcal{F}}x^*_S \cdot \mathbf{1}_{i\in S}$ where $\mathbf{1}_{i\in S}=1$ if $i\in S$ and  $\mathbf{1}_{i\in S}=0$ otherwise. It is easy to verify that  the value of $\hat{y}_i$ represents the probability of item $i$ being selected according to the distribution defined by $x^*$. We next show that to prove this lemma, it suffices to prove that
\begin{eqnarray}
\label{eq:11}
\hat{y}\in B.
\end{eqnarray} As established in \cite{agrawal2010correlation},  if $f$ is monotone and submodular and $\hat{y}_i=\sum_{S\in \mathcal{F}}x^*_S \cdot \mathbf{1}_{i\in S}$, then $ (1-1/e) \sum_{S\in \mathcal{F}}x^*_S f(S) \leq F(\hat{y})$. Here $1-1/e$ is also known as \emph{correlation gap} of monotone submodular functions. Suppose (\ref{eq:11}) is true and  $y^*$ is the optimal solution of $\textbf{P.1}$, it holds that $F(\hat{y})\leq F(y^*)$. Therefore, this lemma is a direct consequence of the observation that $ (1-1/e) \sum_{S\in \mathcal{F}}x^*_S f(S) \leq F(\hat{y}) \leq F(y^*)$.

The rest of the proof is devoted to proving $\hat{y}\in B$. First, because $x^*$ is a feasible solution of  $\textbf{P.0}$, it holds that $\alpha_t\leq \sum_{S\in \mathcal{F}}(x^*_S\cdot |S \cap V_t|) \leq \beta_t, \forall t \in[m]$. It follows that $\alpha_t\leq \sum_{i\in V_t} \hat{y}_i \leq \beta_t, \forall t \in[m]$, this is because $\sum_{S\in \mathcal{F}}(x^*_S\cdot |S \cap V_t|)=\sum_{i\in V_t} \hat{y}_i $ represents the expected number of items being selected from group $V_t$ according to the distribution defined by $x^*$. Second, because $x^*$ is a feasible solution of  $\textbf{P.0}$, the expected number of selected items according to the distribution defined by $x^*$ is at most $b$. Hence, $\sum_{t\in [m]} \sum_{i\in V_t} \hat{y}_i \leq b$. Third, it is trivial to show that $0\leq \hat{y}_i \leq 1, \forall i\in V$. This finishes the proof of $\hat{y}\in B$. $\Box$

\subsection{Algorithm Design}
We next present our algorithm. Initially, we use a continuous greedy algorithm to compute a fractional solution  for $\textbf{P.1}$, which we then round to obtain an integral solution.

\begin{algorithm}[h]
{\small
\caption{Continuous Greedy Algorithm}
\label{alg:greedy-peak}
\begin{algorithmic}[1]
\STATE Set $\delta=9n^2, l=0, y^0=[0]^{n}$.
\WHILE{$l<\delta$}
\STATE For each $i\in V$, estimate $F(i\mid y^l)$
\STATE Find an optimal solution $z \in [0,1]^n $ to $\textbf{P.A}$
\STATE
\begin{center}
\framebox[0.8\textwidth][c]{
\enspace
\begin{minipage}[t]{0.8\textwidth}
\small
\textbf{P.A}
\emph{Maximize$_{y}$ $\sum_{i\in V}y_i F(i\mid y^l)$ }\\
\textbf{subject to:} $y\in B$.
\end{minipage}
}
\end{center}
\vspace{0.1in}
\STATE $y^{l+1}=y^l+z$ \label{line:1}
\STATE Increment $l=l+1$
\ENDWHILE
\STATE $y'\leftarrow y^{\delta}$
\RETURN  $y'$
\end{algorithmic}
}
\end{algorithm}

\textbf{Continuous greedy algorithm.} We first provide a detailed description of the continuous greedy algorithm  (listed in Algorithm \ref{alg:greedy-peak}). The framework of this algorithm was first developed in \cite{calinescu2007maximizing} and we adapt it to find a fractional solution within polytope $B$ (listed in (\ref{eq:2})). Note that polytope $B$ is not downward-closed, which presents unique challenges in our study. This algorithm maintains a fractional solution $y^l\in [0,1]^n$, starting with $y^0=(0,0,\cdots,0)$. In each round $l$, it computes the marginal utility of  each item $i\in V$ on top of $y^l$ with respect to $F$ as follows,
\begin{equation}
\label{eq:1}
F(i\mid y^l)=F(\mathbf{e}_i \vee y^l) - F(y^l).
 \end{equation}
 where $\mathbf{e}_i\in\{0,1\}^n$  is the vector with $1$ in the $i$-th coordinate and $0$ elsewhere; $\mathbf{e}_i \vee y^l$ denotes the element-wise maximum of two vectors $\mathbf{e}_i$ and $y^l$.

 Then we solve the following linear programming problem \textbf{P.A} which assigns a weight $F(i\mid y^l)$ to each item $i$ and seeks the maximum weighted vector in $B$.
\begin{center}
\framebox[0.7\textwidth][c]{
\enspace
\begin{minipage}[t]{0.7\textwidth}
\small
\textbf{P.A}
\emph{Maximize$_{y}$ $\sum_{i\in V}y_i F(i\mid y^l)$ } \textbf{subject to:} $y\in B$.
\end{minipage}
}
\end{center}
\vspace{0.1in}
\vspace{0.1in}

After solving \textbf{P.A} at round $l$ and obtaining  an optimal solution $z\in [0,1]^n$, we update the fractional solution as follows: $y^{l+1}=y^l+z$.  After $\delta$ rounds where $\delta=9n^2$,  $ y^{\delta}$ is returned as the final solution $y'$.

\textbf{Rounding.} We next employ  pipage rounding \cite{ageev2004pipage},  a simple deterministic procedure of rounding of linear relaxations, to round $y'$ to an integral solution. This algorithm is composed of three phases.
\begin{itemize}
\item Phase 1: For each $t\in[m]$, repeatedly perform the following until $V_t$ has no more than one non-integral coordinate: Choose any two fractional coordinates $i$, $j$ such that $i, j\in V_t$. Calculate $\theta_1=\min\{1-y'_i, y'_j\}$ and $\theta_2=\min\{y'_i, 1-y'_j\}$. Create two vectors, $y^a=y'+\theta_1(\mathbf{e}_i-\mathbf{e}_j)$ and $y^b=y'+\theta_2(\mathbf{e}_j-\mathbf{e}_i)$. If $F(y^a)\geq F(y^b)$, set $y\leftarrow y_a$, otherwise set $y\leftarrow y^b$.
\item Phase 2: Assume $y_1, \cdots , y_k$ are the remaining fractional coordinates. Repeat the same procedure as in the first phrase until $y$ has at most one non-integral coordinate.
\item  Phase 3: Let $i$ denote the last non-integral coordinate, if any. Set $y_i=1$. Output $A\subseteq V$ whose coordinate  in $y$ is $1$.
\end{itemize}

Note that a similar framework has been utilized to tackle the fair submodular maximization problem in a deterministic setting \cite{celis2018multiwinner}. This problem aims to identify a fixed set of items that optimize a submodular function while fulfilling group fairness constraints. Their approach shares similarities with ours in the rounding stage, but does not require the third phase. This is because in their setting, both $\alpha_t$ and $\beta_t$ are integers, which allows them to ensure that no non-integral coordinates exist after the first two rounding phases.

\subsection{Performance Analysis}
Recall that $x^*$ denotes the optimal solution of $\textbf{P.0}$, let $OPT=\sum_{S\in \mathcal{F}}x^*_S f(S)$ denote the utility of the optimal solution. The following theorem states that $A$, the solution set returned from our algorithm, is a \emph{near} feasible solution of $\textbf{P.0}$ and has a utility of at least $(1-1/e)^2 OPT$.
\begin{theorem}
\label{thm:1}
Let $A$ be the set returned by our algorithm and $OPT$ be the utility of the optimal solution of $\textbf{P.0}$. It follows that:
\begin{eqnarray}
f(A)\geq (1-1/e)^2 OPT.
\end{eqnarray}
Moreover, $A$ always satisfies the cardinality constraint and nearly satisfies the fairness constraints of $\textbf{P.0}$, i.e.,   $|A|\leq b$ and $\lfloor \alpha_t\rfloor \leq |A \cap V_t| \leq \lceil\beta_t\rceil, \forall t \in[m]$.
\end{theorem}
\emph{Proof:} We first prove that $|A|\leq b$ always holds. Observe that the fractional solution $y'$ found by the continuous greedy algorithm belongs to $B$, hence, $\sum_{i\in V} y'_i \leq b$. Moreover, phases 1 and 2 in the rounding stage do not change this value, and phase 3 rounds the last non-integral coordinate, if any, to one. It follows that $|A|\leq \lceil\sum_{i\in V} y'_i\rceil \leq b$ where the second inequality is by the observations that $\sum_{i\in V} y'_i \leq b$ and $b$ is an integer.

We next prove that $A$ nearly satisfies the fairness constraints of $\textbf{P.0}$, i.e.,  $\lfloor \alpha_t\rfloor \leq |A \cap V_t|  \leq \lceil\beta_t\rceil, \forall t \in[m]$. Because $y'\in B$, it holds that $\alpha_t\leq \sum_{i\in V_t} y'_i \leq \beta_t, \forall t \in[m]$. Observe that  phase 1 does not change this value, phases 2 and 3 round at most one fractional coordinate from each group to a binary value. Hence, $\lfloor \sum_{i\in V_t} y'_i\rfloor \leq |A \cap V_t|  \leq \lceil\sum_{i\in V_t} y'_i\rceil, \forall t \in[m]$. This, together with $\alpha_t\leq \sum_{i\in V_t} y'_i \leq \beta_t, \forall t \in[m]$, implies that  $\lfloor \alpha_t\rfloor \leq |A \cap V_t|  \leq \lceil\beta_t\rceil, \forall t \in[m]$.

At last, we prove the approximation ratio of $A$. Recall that $y^*$ denotes the optimal solution of $\textbf{P.1}$, \cite{calinescu2007maximizing} has proved that if $f$ is monotone and submodular, then the fractional solution $y'$ returned from the continuous greedy algorithm has a utility of at least $(1-1/e) F(y^*)$, i.e.,   $F(y')\geq (1-1/e) F(y^*)$. This, together with Lemma \ref{lem:1}, implies that $F(y')\geq (1-1/e)^2 \sum_{S\in \mathcal{F}}x^*_S f(S)= (1-1/e)^2 OPT$. To prove $f(A)\geq (1-1/e)^2 OPT$, it suffices to show that $f(A)\geq F(y')$. We next prove this inequality. Observe that in phases 1 and 2 of the rounding stage, we perform pipage rounding to round $y'$ to a vector $y$ that contains at most one non-integral coordinate. According to \cite{calinescu2007maximizing}, pipage rounding does not decrease the expected utility of $y'$, that is, $F(y)\geq F(y')$. In phase 3, we round the last non-integral coordinate in $y$, if any, to one. This operation does not decrease the expected utility of $y$ by the assumption that $f$ is monotone. Hence, $F(y)\geq F(y')$ still holds. Recall that $y$ is the indicator vector of $A$, hence, $f(A)=F(y)$. Therefore, $f(A)=F(y)\geq F(y')$. $\Box$

\textbf{Remark 1:} It follows immediately from the preceding theorem that if $\alpha_t$ and $\beta_t$ are both integers for all $t\in[m]$, then our solution strictly satisfies all fairness constraints of problem $\textbf{P.0}$.

\subsection{A Fast Greedy Algorithm}
Our prior  algorithm involves solving a multi-linear relaxation problem, which can be slow and computationally expensive, particularly for large scale problems. In this section, we introduce a simple greedy algorithm that offers a significant increase in speed but with a trade-off in the form of a decreased approximation ratio.

Even though $\textbf{P.0}$ permits the use of randomized solutions, Theorem \ref{thm:1} shows that a deterministic solution is sufficient for obtaining a constant-factor approximation for $\textbf{P.0}$. We next present a simple greedy algorithm that effectively finds a near optimal deterministic solution, which in turn results in a constant-factor approximation for the problem $\textbf{P.0}$. To this end we introduce a new optimization problem $\textbf{P.2}$, a deterministic version of $\textbf{P.0}$ (with relaxed fairness constraints).

 \begin{center}
\framebox[0.4\textwidth][c]{
\enspace
\begin{minipage}[t]{0.4\textwidth}
\small
$\textbf{P.2}$
$\max_{S\in \mathcal{F}} f(S)$ \\
\textbf{subject to:}  \\
$\lfloor\alpha_t\rfloor\leq |S \cap V_t| \leq \lceil\beta_t\rceil, \forall t \in[m]$.
\end{minipage}
}
\end{center}
\vspace{0.1in}

Note that in  $\textbf{P.2}$ we use $\lfloor\alpha_t\rfloor$ and $\lceil\beta_t\rceil$ as lower and upper bounds, hence a feasible solution of $\textbf{P.2}$ is a near feasible solution of the original problem $\textbf{P.0}$. The following lemma states that the optimal solution of  $\textbf{P.2}$ attains a $(1-1/e)^2$ approximation of the problem $\textbf{P.0}$.
\begin{lemma}
\label{lem:2}
Let $A^{P2}$ denote the optimal solution of $\textbf{P.2}$, it holds that $f(A^{P2})\geq(1-1/e)^2 OPT$ where $OPT$ is the optimal solution of $\textbf{P.0}$.
\end{lemma}
\emph{Proof:} Recall that in  Theorem \ref{thm:1}, we show that $f(A)\geq(1-1/e)^2 OPT$ where $A$ satisfies all constraints in $\textbf{P.2}$. Because $A^{P2}$ is the optimal solution of $\textbf{P.2}$, we have $f(A^{P2})\geq f(A)\geq(1-1/e)^2 OPT$. $\Box$

We next present a simple greedy algorithm to attain a $1/2$ approximation of $\textbf{P.2}$. First, we present $\textbf{P.3}$, a relaxed problem of $\textbf{P.2}$.
 \begin{center}
\framebox[0.4\textwidth][c]{
\enspace
\begin{minipage}[t]{0.4\textwidth}
\small
$\textbf{P.3}$
$\max_{S\subseteq V} f(S)$ \\
\textbf{subject to:}
\begin{equation*}
\begin{cases}
|S \cap V_t| \leq \lceil\beta_t\rceil, \forall t \in[m].\\
\sum_{t\in [m]} \max\{\lfloor\alpha_t\rfloor, |S \cap V_t|\} \leq b.
\end{cases}
\end{equation*}
\end{minipage}
}
\end{center}
\vspace{0.1in}

It is easy to verify that any feasible solution of $\textbf{P.2}$ must be a feasible solution of $\textbf{P.3}$. Hence, $f(A^{P2})\leq f(A^{P3})$ where $A^{P3}$ is the optimal solution of $\textbf{P.3}$.  It has been shown that the constraints listed in $\textbf{P.3}$ constitute a matroid \cite{el2020fairness}. Hence, $\textbf{P.3}$ is a classic submodular maximization problem subject to a matroid constraint. A simple greedy algorithm guarantees  a $1/2$ approximation of $\textbf{P.3}$ \cite{nemhauser1978analysis}. This algorithm works by iteratively adding items to the solution set such that at each step, the marginal increase in the objective value is maximized, and the matroid constraint is satisfied, and it terminates when the current solution set can not be expanded. Let $A^g$ denote the solution returned from the greedy algorithm, it holds that
\begin{eqnarray}
\label{eq:3}
f(A^g)\geq (1/2)  f(A^{P3}) \geq (1/2) f(A^{P2}).
\end{eqnarray}

 Moreover, it is easy to verify that $A^g$ must be a feasible solution of $\textbf{P.2}$.

\begin{theorem}
Let $A^g$ denote the solution of returned from the greedy algorithm, it holds that $f(A^g)\geq\frac{(1-1/e)^2}{2} \cdot OPT$. Moreover, $A^g$  always satisfies the cardinality constraint and nearly satisfies the fairness constraints of $\textbf{P.0}$, i.e.,   $|A^g|\leq b$ and $\lfloor \alpha_t\rfloor \leq |A^g \cap V_t| \leq \lceil\beta_t\rceil, \forall t \in[m]$.
\end{theorem}
\emph{Proof:} The proof of the first part of this theorem stems from inequality (\ref{eq:3}) and Lemma \ref{lem:2}. The second part of this theorem is because $A^g$ is a feasible solution to problem $\textbf{P.2}$. $\Box$

\section{A Feasible $(1-1/e)$-approximation Randomized Algorithm}
\label{sec:2}
We now present a randomized algorithm for $\textbf{P.0}$. In contrast to the results presented in the previous section, our randomized solution offers three advantages: (1) our solution does not rely on the assumption of  non-overlapping groups, (2) our solution satisfies all fairness constraints in a \emph{strict} sense, and (3) we achieve  the optimal approximation ratio of $1-1/e$.


 As previously stated, $\textbf{P.0}$ has a number of variables equal to the number of elements in $\mathcal{F}$, which can become extremely large when $n$ is significant. This means that standard LP solvers are unable to handle this LP problem effectively. To tackle this issue, we resort to its corresponding dual problem ($\textbf{Dual of P.0}$) and utilize the ellipsoid algorithm \cite{grotschel1981ellipsoid} to solve it. In the dual problem, we assign two
``weights'' $z_t\in \mathbb{R}_{\geq 0}$ and $u_t\in \mathbb{R}_{\geq 0}$ to each group $V_t$ and there is an additional variable $w\in \mathbb{R}_{\geq 0}$.

  \begin{center}
   \framebox[0.8\textwidth][c]{
\enspace
\begin{minipage}[t]{0.8\textwidth}
\small
$\textbf{Dual of P.0}$
$\min_{z\in \mathbb{R}_{\geq 0}^m, u\in \mathbb{R}_{\geq 0}^m, w\in \mathbb{R}_{\geq 0}}  \sum_{t\in [m]}(\beta_tu_t -\alpha_t z_t)+w$\\
\textbf{subject to:}
$w \geq f(S)+\sum_{t\in [m]}  |S \cap V_t| \cdot (z_t-u_t), \forall S\in \mathcal{F}. $
\end{minipage}
}
\end{center}
\vspace{0.1in}

At a high level, the ellipsoid method is a powerful tool used to determine the emptiness of a given non-degenerate convex set $C$. For our problem,  $C$ represents the feasible region of $\textbf{Dual of P.0}$. This method begins by defining an ellipsoid that is guaranteed to contain $C$. In each iteration, the algorithm checks if the center of the current ellipsoid is located within $C$. If it is, this means that $C$ is nonempty and that the current solution is feasible. In this case, the method tries a smaller objective. On the other hand, if the center of the current ellipsoid is not located within $C$, the method uses an (approximate) separation oracle to identify a violated constraint and finds a smaller ellipsoid with a center that satisfies that constraint. This process continues until a feasible solution is found or the volume of the bounding ellipsoid is so small that $C$ is considered empty, meaning that no feasible solution with a smaller objective can be found. It is important to note that this method operates in a geometric fashion and does not require an explicit description of the LP. The only requirement is a polynomial-time (approximate) separation oracle, which can determine whether a point lies within $C$ and, if not, return a separating hyperplane. This method has a polynomial number of iterations for solving linear problems.

In the context of our problem, we approximately solve the \textsc{SubMax} problem to check the feasibility of the current solution and act as the separation oracle.
\begin{definition}[SubMax]
\label{def:min}
Given a utility function $f$, a cardinality constraint $b$, and two vectors $z\in \mathbb{R}_{\geq 0}^m$ and $u\in \mathbb{R}_{\geq 0}^m$,  \textsc{SubMax}$(z, u, b)$ aims to
\begin{eqnarray}
\max_{S\in \mathcal{F}} (f(S)+\sum_{t\in [m]}  |S \cap V_t| \cdot (z_t-u_t)).
\end{eqnarray}
\end{definition}

   \textsc{SubMax}$(z, u, b)$ asks for a set $S$ of size at most $b$ such that $f(S)+\sum_{t\in [m]}  |S \cap V_t| \cdot (z_t-u_t)$ is maximized. Observe that $f$ is non-negative monotone and submodular; and  $\sum_{t\in [m]}  |S \cap V_t| \cdot (z_t-u_t)$ is a modular function in terms of $S$, hence,    \textsc{SubMax}$(z, u, b)$ is a classic problem of maximizing the summation of a non-negative  monotone submodular and a modular function under cardinality constraints. \cite{sviridenko2017optimal} presented a randomized polynomial time algorithm that produces a set $A$ such that for every $S\in \mathcal{F}$, it holds that
\begin{eqnarray}
\label{eq:haha}
f(A)+\sum_{t\in [m]}  |A \cap V_t| \cdot  (z_t-u_t) \geq (1-1/e) f(S)+\sum_{t\in [m]}  |S \cap V_t| \cdot (z_t-u_t).
\end{eqnarray}

Now we are ready to present the main theorem of this section.

\begin{theorem}
\label{thm:2}
There exists a polynomial time $(1-1/e)$-approximation algorithm (with additive error $\epsilon$) for $\textbf{P.0}$.
\end{theorem}

The rest of this section is devoted to proving Theorem \ref{thm:2}, that is, we present a polynomial $(1-1/e)$-approximation algorithm for $\textbf{P.0}$. Let $C(L)$ denote the set of $(z\in \mathbb{R}_{\geq 0}^m, u\in \mathbb{R}_{\geq 0}^m, w\in \mathbb{R}_{\geq 0})$ satisfying that
\[\sum_{t\in [m]}(\beta_tu_t -\alpha_t z_t)+w \leq L,\]
\[w \geq f(S)+\sum_{t\in [m]}  |S \cap V_t| \cdot (z_t-u_t), \forall S\in \mathcal{F}.\]

We use binary search to determine the smallest value of $L$ for which $C(L)$ is not empty. For a given $L$, we first evaluate the inequality $\sum_{t\in [m]}(\beta_tu_t -\alpha_t z_t)+w \leq L$. Then, we run algorithm from \cite{sviridenko2017optimal} (labeled as  $\mathcal{A}$) to solve \textsc{SubMax}$(z, u, b)$. 
Let $A$ be the solution set returned from $\mathcal{A}$.
\begin{itemize}
\item If  $f(A)+\sum_{t\in [m]}  |A \cap V_t| \cdot (z_t-u_t) \leq w$, then $C(L)$ is marked as non-empty. In this case, we try a smaller $L$.
\item If $f(A)+\sum_{t\in [m]}  |A \cap V_t| \cdot (z_t-u_t) > w$, then $(z, w)\notin C(L)$ and $A$ is a separating hyperplane. We find a smaller ellipsoid with a center that satisfies that constraint. This process continues until a feasible solution in $C(L)$ is found (in this case, we try a smaller $L$) or the volume of the bounding ellipsoid is so small that $C(L)$ is considered empty (in this case, we conclude that the current objective is not achievable and will therefore try a larger $L$).
\end{itemize}

To learn about the specific steps involved in running ellipsoid using separation oracles and achieving (multiplicative and additive) approximate guarantees, we suggest referring to Chapter 2 of \cite{bubeck2015convex}. Assume $L^*$ is the minimum value of $L$ for which the algorithm determines that $C(L)$ is non-empty. Hence, there exists a  $(z^*, u^*, w^*)$ such that
\begin{eqnarray}
\label{eq:pp}
\sum_{t\in [m]}(\beta_tu^*_t -\alpha_t z^*_t)+w^* \leq L^*
\end{eqnarray} and
\begin{eqnarray}
\label{eq:oo}
f(A)+\sum_{t\in [m]}  |A \cap V_t| \cdot (z^*_t-u^*_t) \leq w^*
\end{eqnarray}
 where $A$ is the output obtained from algorithm $\mathcal{A}$ after solving \textsc{SubMax}$(z^*, u^*, b)$. Let $\mu=1-1/e$, it follows that $ \forall S\in \mathcal{F},$
\begin{eqnarray}
\label{eq:6}
 f(S)+\sum_{t\in [m]}  |S \cap V_t| \cdot (u^*_t-z^*_t)/\mu &&\leq (f(A)+\sum_{t\in [m]}  |A \cap V_t| \cdot(u^*_t-z^*_t))/\mu ~\nonumber \\
 && \leq w^*/\mu
 \end{eqnarray}
 where the first inequality is by (\ref{eq:haha}) and the second inequality is by inequality (\ref{eq:oo}). In addition, inequality (\ref{eq:pp}) implies that
 \begin{eqnarray}
 \label{eq:7}
 \sum_{t\in [m]} (\beta_tu^*_t -\alpha_t z^*_t)/\mu+w^*/\mu \leq L^*/\mu.
 \end{eqnarray}

Inequality (\ref{eq:6}) implies that $(z^*/\mu, u^*/\mu, w^*/\mu)$ is a feasible solution of  $\textbf{Dual of P.0}$. This, together with inequality (\ref{eq:7}), implies that the optimal solution of $\textbf{Dual of P.0}$ and thus the optimal solution of $\textbf{P.0}$ (by strong duality) is  upper bounded by $\frac{1}{\mu} \cdot L^*$. By finding a solution to $\textbf{P.0}$ with a value of $L^*$, we attain a $\mu$-approximation solution for the original problem $\textbf{P.0}$. Here, we explain how to  compute such a  solution  using only feasible solution sets  corresponding to the separating hyperplanes found by the separation oracle. Assume $L^*-\epsilon$ is the largest value of $L$ for which the algorithm determines that $C(L)$ is empty, where $\epsilon$ is decided by the precision of our algorithm. Let $\mathcal{F}'$ denote the subset of $\mathcal{F}$  for which the dual constraint is violated
in the implementation of the ellipsoid algorithm on $C (L^* - \epsilon)$.
Then, $|\mathcal{F}'|$ is polynomial. We consider the following polynomial sized  $\textbf{Dual of P.0}$ (labeled as $\textbf{Poly-sized Dual of P.0}$), using separating hyperplanes from $\mathcal{F}'$.

  \begin{center}
   \framebox[0.9\textwidth][c]{
\enspace
\begin{minipage}[t]{0.9\textwidth}
\small
$\textbf{Poly-sized Dual of P.0}$
$\min_{z\in \mathbb{R}_{\geq 0}^m, u\in \mathbb{R}_{\geq 0}^m, w\in \mathbb{R}_{\geq 0}}  \sum_{t\in [m]}(\beta_tu_t -\alpha_t z_t)+w$\\
\textbf{subject to:}
$w \geq f(S)+\sum_{t\in [m]}  |S \cap V_t| \cdot (z_t-u_t), \forall S\in \mathcal{F'}. $
\end{minipage}
}
\end{center}
\vspace{0.1in}

Because $C(L^* - \epsilon)$ is empty, the value of the optimal solution to $\textbf{Poly-sized Dual of P.0}$ is at least $L^*-\epsilon$. Hence, the value of
 the dual of $\textbf{Poly-sized Dual of P.0}$, which is listed in $\textbf{Poly-sized P.0}$, is at least  $L^*-\epsilon$. Note that $\textbf{Poly-sized P.0}$ is of polynomial size.
 \begin{center}
\framebox[0.6\textwidth][c]{
\enspace
\begin{minipage}[t]{0.6\textwidth}
\small
$\textbf{Poly-sized P.0}$
$\max_{x\in [0,1]^{\mathcal{F}'}} \sum_{S\in \mathcal{F}'}x_S f(S)$ \\
\textbf{subject to:}
\begin{equation*}
\begin{cases}
\alpha_t\leq \sum_{S\in \mathcal{F}}(x_S\cdot |S \cap V_t|) \leq \beta_t, \forall t \in[m].\\
\sum_{S\in \mathcal{F}'}x_S\leq 1.
\end{cases}
\end{equation*}
\end{minipage}
}
\end{center}
\vspace{0.1in}

Recall that the optimal solution of $\textbf{P.0}$ is  upper bounded by $\frac{1}{\mu} \cdot L^*$, obtaining an optimal solution from $\textbf{Poly-sized P.0}$ provides a $\mu$-approximation for $\textbf{P.0}$ (with additive error $\epsilon$), where $\mu=1-1/e$.

\section{Conclusion}
We introduce a set of fairness criteria, including restricted dominance and minority protection, to ensure that the proportion of selected items from each group falls within the desired range. We present a $(1-1/e)^2$-approximation deterministic algorithm and a fast greedy deterministic algorithm that approximately satisfy the fairness constraints. Additionally, we propose a randomized algorithm that strictly satisfies all fairness constraints and achieves the optimal approximation ratio of $1-1/e$.

\bibliographystyle{splncs04}
\bibliography{reference}




\end{document}